\documentstyle[aps,graphicx]{revtex}

\def\beq{\begin{equation}}
\def\eeq{\end{equation}}
\def\beqa{\begin{eqnarray}}
\def\eeqa{\end{eqnarray}}

\def\za{\alpha}
\def\zb{\beta}
\def\lsim{\mathrel{\raise.3ex\hbox{$<$\kern-.75em\lower1ex\hbox{$\sim$}}} }
\def\gsim{\mathrel{\raise.3ex\hbox{$>$\kern-.75em\lower1ex\hbox{$\sim$}}} }

\begin{document}
\addtolength{\baselineskip}{.2cm}
\onecolumn

\begin{flushright}
NCU-HEP-k006\\
Sep 2002
\end{flushright}

\vspace*{1in}

\begin{center}
{\Large \bf Fermion Dipole Moments from R-parity Violating Parameters.$^\star$}

\vspace*{.5in}
{\bf  Otto C.W. Kong}\\[.08in]
{\it Department of Physics, National Central University, Chung-li, TAIWAN 32054}

\vspace*{1.5in}
{Abstract}\\
\end{center} 
We have developed an efficient formulation for the study of the generic supersymmetric
standard model, which admits all kind of R-parity violating terms. Using the formulation, 
we discuss all sources of fermion dipole moment contributions from
R-parity violating, or rather lepton number violating,  parametersand the constraints 
obtained. Stringent constraints comparable to those from neutrino masses are resulted
in some cases.

\noindent

\vfill
\noindent --------------- \\
$^\star$ Talk presented  at SUSY'02 conference (Jun 17-23), DESY, Hamburg, Germany. 
\\
 --- submission for the proceedings.  
 
\clearpage

\title{Fermion Dipole Moments from R-parity Violating Parameters.}

\author{Otto C. W. Kong}

\address{Department of Physics, National Central University, Chung-li, TAIWAN 32054
\\E-mail: otto@phy.ncu.edu.tw}
\maketitle

\begin{abstract}
We have developed an efficient formulation for the study of the generic supersymmetric
standard model, which admits all kind of R-parity violating terms. Using the formulation, 
we discuss all sources of fermion dipole moment contributions from
R-parity violating, or rather lepton number violating,  parametersand the constraints 
obtained. Stringent constraints comparable to those from neutrino masses are resulted
in some cases.
\end{abstract}
\thispagestyle{empty}

\section{Introduction}
Fermion electric dipole moments (EDMs) are known to be extremely useful 
constraints on (the CP violating part of) models depicting interesting 
scenarios of beyond Standard Model (SM) physics. In particular, the 
experimental bounds on neutron EDM ($d_n$) and electron EDM ($d_e$) are very 
stringent. The current numbers are given by 
$d_n < 6.3 \cdot 10^{-26}\,e \cdot \mbox{cm}$
and $d_e < 4.3 \cdot 10^{-27}\,e \cdot \mbox{cm}$. The SM contributions are 
known to be very small, given that the only source of CP violation has to 
come from the KM phase in (charged current) quark flavor mixings :
  $d_n \sim 10^{-32}\,e\cdot \mbox{cm}$ and 
$d_e \sim 8 \cdot 10^{-41}\,e\cdot \mbox{cm}$. 

Extensions of the SM normally are expected to have potentially large EDM 
contributions. For instance, for the minimal supersymmetric standard model (MSSM),
there are a few source of such new contributions. For example, they can come in 
through $LR$ sfermion mixings. The latter have two parts, an $A$-term contribution 
as well as a $F$-term contribution. The $F$-term is a result of the complex phase
in the so-called $\mu$-term. The resulted constraints on MSSM have been studied
extensively. We are interested here in the modified version with R parity 
not imposed. We will illustrate that there are extra contributions at the same 
level and discuss the class of important constraints hence resulted.

\section{Formulation and Notation}
A theory built with the minimal superfield spectrum incorporating the SM particles, 
the admissible renormalizable interactions dictated by the SM (gauge) symmetries 
together with the idea that supersymmetry (SUSY) is softly broken is what should 
be called the the generic supersymmetric standard model (GSSM). The popular
MSSM differs from the generic version in having a discrete symmetry, called R 
parity, imposed by hand to enforce baryon and lepton number conservation. With 
the strong experimental hints at the existence of lepton number violating neutrino 
masses, such a theory of SUSY without R-parity deserves ever more attention. The
GSSM contains all kinds of (so-called) R-parity violating (RPV) parameters.
The latter includes the more popular trilinear ($\lambda_{ijk}$, $\lambda_{ijk}^{\prime}$, and	$\lambda_{ijk}^{\prime\prime}$) and bilinear 
($\mu_i$) couplings in the superpotential, as well as  soft SUSY breaking
parameters of the trilinear, bilinear, and soft mass (mixing) types. In order not 
to miss any plausible RPV phenomenological features, it is important that all of 
the RPV parameters be taken into consideration without {\it a priori} bias. 
We do, however, expect some sort of symmetry principle to guard against the very 
dangerous proton decay problem. The emphasis is hence put on the lepton number
violating phenomenology.

The renormalizable superpotential for the GSSM can be written  as
\small\beqa
W \!\! &=& \!\varepsilon_{ab}\Big[ \mu_{\alpha}  \hat{H}_u^a \hat{L}_{\alpha}^b 
+ h_{ik}^u \hat{Q}_i^a   \hat{H}_{u}^b \hat{U}_k^{\scriptscriptstyle C}
+ \lambda_{\alpha jk}^{\!\prime}  \hat{L}_{\alpha}^a \hat{Q}_j^b
\hat{D}_k^{\scriptscriptstyle C} 
+
\frac{1}{2}\, \lambda_{\alpha \beta k}  \hat{L}_{\alpha}^a  
 \hat{L}_{\beta}^b \hat{E}_k^{\scriptscriptstyle C} \Big] + 
\frac{1}{2}\, \lambda_{ijk}^{\!\prime\prime}  
\hat{U}_i^{\scriptscriptstyle C} \hat{D}_j^{\scriptscriptstyle C}  
\hat{D}_k^{\scriptscriptstyle C}   ,
\eeqa\normalsize
where  $(a,b)$ are $SU(2)$ indices, $(i,j,k)$ are the usual family (flavor) 
indices, and $(\za, \zb)$ are extended flavor indices going from $0$ to $3$.
At the limit where $\lambda_{ijk}, \lambda^{\!\prime}_{ijk},  
\lambda^{\!\prime\prime}_{ijk}$ and $\mu_{i}$  all vanish, 
one recovers the expression for the R-parity preserving MSSM, 
with $\hat{L}_{0}$ identified as $\hat{H}_d$. Without R-parity imposed,
the latter is not {\it a priori} distinguishable from the $\hat{L}_{i}$'s.
Note that $\lambda$ is antisymmetric in the first two indices, as
required by  the $SU(2)$  product rules, as shown explicitly here with 
$\varepsilon_{\scriptscriptstyle 12} =-\varepsilon_{\scriptscriptstyle 21}=1$.
Similarly, $\lambda^{\!\prime\prime}$ is antisymmetric in the last two 
indices, from $SU(3)_{\scriptscriptstyle C}$. 

R-parity is exactly an {\it ad hoc} symmetry put in to make $\hat{L}_{0}$,
stand out from the other $\hat{L}_i$'s as the candidate for  $\hat{H}_d$.
It is defined in terms of baryon number, lepton number, and spin as, 
explicitly, ${\mathcal R} = (-1)^{3B+L+2S}$. The consequence is that 
the accidental symmetries of baryon number and lepton number in the SM 
are preserved, at the expense of making particles and superparticles having 
a categorically different quantum number, R parity. The latter is actually 
not the most effective discrete symmetry to control superparticle 
mediated proton decay\cite{pd}, but is most restrictive in terms
of what is admitted in the Lagrangian, or the superpotential alone. 
On the other hand, R parity also forbides neutrino masses in the
supersymmetric SM. The strong experimental hints for the existence of 
(Majorana) neutrino masses is an indication of lepton 
number violation, hence suggestive of R-parity violation.

\thispagestyle{empty}

The soft SUSY breaking part 
of the Lagrangian is more interesting, if only for the fact that  many
of its interesting details have been overlooked in the literature.
However, we will postpone the discussion till after we address the
parametrization issue.

Doing phenomenological studies without specifying a choice 
of flavor bases is ambiguous. It is like doing SM quark physics with 18
complex Yukawa couplings, instead of the 10 real physical parameters.
As far as the SM itself is concerned, the extra 26 real parameters
are simply redundant, and attempts to relate the full 36 parameters to
experimental data will be futile. In the GSSM, the choice of an optimal
parametrization mainly concerns the 4 $\hat{L}_\alpha$ flavors. We use
here the single-VEV parametrization\cite{ru,as8} (SVP), in which flavor 
bases are chosen such that : 
1/ among the $\hat{L}_\alpha$'s, only  $\hat{L}_0$, bears a VEV,
{\it i.e.} {\small $\langle \hat{L}_i \rangle \equiv 0$};
2/  {\small $h^{e}_{jk} (\equiv \lambda_{0jk}) 
=\frac{\sqrt{2}}{v_{\scriptscriptstyle 0}} \,{\rm diag}
\{m_{\scriptscriptstyle 1},
m_{\scriptscriptstyle 2},m_{\scriptscriptstyle 3}\}$};
3/ {\small $h^{d}_{jk} (\equiv \lambda^{\!\prime}_{0jk} =-\lambda_{j0k}) 
= \frac{\sqrt{2}}{v_{\scriptscriptstyle 0}}{\rm diag}\{m_d,m_s,m_b\}$}; 
4/ {\small $h^{u}_{ik}=\frac{\sqrt{2}}{v_{\scriptscriptstyle u}}
V_{\mbox{\tiny CKM}}^{\!\scriptscriptstyle T} \,{\rm diag}\{m_u,m_c,m_t\}$}, 
where ${v_{\scriptscriptstyle 0}} \equiv  \sqrt{2}\,\langle \hat{L}_0 \rangle$
and ${v_{\scriptscriptstyle u} } \equiv \sqrt{2}\,
\langle \hat{H}_{u} \rangle$. The big advantage of the SVP is that it gives 
the complete tree-level mass matrices of all the states (scalars and fermions) 
the simplest structure\cite{as5,as8}.

\section{Leptons in GSSM}
The SVP gives quark mass matrices exactly in the SM form. For the masses
of the color-singlet fermions, all the RPV effects are paramatrized by the
$\mu_i$'s only. For example, the five charged fermions ( gaugino
+ Higgsino + 3 charged leptons ), we have
\small\beq \label{mc}
{\mathcal{M}_{\scriptscriptstyle C}} =
 \left(
{\begin{array}{ccccc}
{M_{\scriptscriptstyle 2}} &  
\frac{g_{\scriptscriptstyle 2}{v}_{\scriptscriptstyle 0}}{\sqrt 2}  
& 0 & 0 & 0 \\
 \frac{g_{\scriptscriptstyle 2}{v}_{\scriptscriptstyle u}}{\sqrt 2} & 
 {{ \mu}_{\scriptscriptstyle 0}} & {{ \mu}_{\scriptscriptstyle 1}} &
{{ \mu}_{\scriptscriptstyle 2}}  & {{ \mu}_{\scriptscriptstyle 3}} \\
0 &  0 & {{m}_{\scriptscriptstyle 1}} & 0 & 0 \\
0 & 0 & 0 & {{m}_{\scriptscriptstyle 2}} & 0 \\
0 & 0 & 0 & 0 & {{m}_{\scriptscriptstyle 3}}
\end{array}}
\right)  \; .
\eeq\normalsize
Moreover each $\mu_i$ parameter here characterizes directly the RPV effect
on the corresponding charged lepton  ($\ell_i = e$, $\mu$, and $\tau$).
This, and the corresponding neutrino-neutralino masses and mixings,
has been exploited to implement a detailed study of the tree-level
RPV phenomenology from the gauge interactions, with interesting 
results\cite{ru}.

\thispagestyle{empty}

Neutrino masses and oscillations is no doubt one of the most important aspects
of the model. Here, it is particularly important that the various RPV 
contributions to neutrino masses, up to 1-loop level, be studied in a 
framework that takes no assumption on the other parameters. Our formulation 
provides such a framework. Interested readers are referred to 
Refs.\cite{as5,ok,as1,as9,AL}.

\section{Soft SUSY Breaking Terms and the Scalar Masses}
Obtaining the squark and slepton masses is straightforward, once all the 
admissible soft SUSY breaking terms are explicitly written down\cite{as5}. 
The soft SUSY breaking part of the Lagrangian can be written as 
\beqa
V_{\rm soft} &=& \epsilon_{\!\scriptscriptstyle ab} 
  B_{\za} \,  H_{u}^a \tilde{L}_\za^b +
\epsilon_{\!\scriptscriptstyle ab} \left[ \,
A^{\!\scriptscriptstyle U}_{ij} \, 
\tilde{Q}^a_i H_{u}^b \tilde{U}^{\scriptscriptstyle C}_j 
+ A^{\!\scriptscriptstyle D}_{ij} 
H_{d}^a \tilde{Q}^b_i \tilde{D}^{\scriptscriptstyle C}_j  
+ A^{\!\scriptscriptstyle E}_{ij} 
H_{d}^a \tilde{L}^b_i \tilde{E}^{\scriptscriptstyle C}_j   \,
\right] + {\rm h.c.}\nonumber \\
&+&
\epsilon_{\!\scriptscriptstyle ab} 
\left[ \,  A^{\!\scriptscriptstyle \lambda^\prime}_{ijk} 
\tilde{L}_i^a \tilde{Q}^b_j \tilde{D}^{\scriptscriptstyle C}_k  
+ \frac{1}{2}\, A^{\!\scriptscriptstyle \lambda}_{ijk} 
\tilde{L}_i^a \tilde{L}^b_j \tilde{E}^{\scriptscriptstyle C}_k  
\right] 
+ \frac{1}{2}\, A^{\!\scriptscriptstyle \lambda^{\prime\prime}}_{ijk} 
\tilde{U}^{\scriptscriptstyle C}_i  \tilde{D}^{\scriptscriptstyle C}_j  
\tilde{D}^{\scriptscriptstyle C}_k  + {\rm h.c.}
\nonumber \\
&+&
 \tilde{Q}^\dagger \tilde{m}_{\!\scriptscriptstyle {Q}}^2 \,\tilde{Q} 
+\tilde{U}^{\dagger} 
\tilde{m}_{\!\scriptscriptstyle {U}}^2 \, \tilde{U} 
+\tilde{D}^{\dagger} \tilde{m}_{\!\scriptscriptstyle {D}}^2 
\, \tilde{D} 
+ \tilde{L}^\dagger \tilde{m}_{\!\scriptscriptstyle {L}}^2  \tilde{L}  
  +\tilde{E}^{\dagger} \tilde{m}_{\!\scriptscriptstyle {E}}^2 
\, \tilde{E}
+ \tilde{m}_{\!\scriptscriptstyle H_{\!\scriptscriptstyle u}}^2 \,
|H_{u}|^2 
\nonumber \\
&& + \frac{M_{\!\scriptscriptstyle 1}}{2} \tilde{B}\tilde{B}
   + \frac{M_{\!\scriptscriptstyle 2}}{2} \tilde{W}\tilde{W}
   + \frac{M_{\!\scriptscriptstyle 3}}{2} \tilde{g}\tilde{g}
+ {\rm h.c.}\; ,
\label{soft}
\eeqa
where we have separated the R-parity conserving $A$-terms from the 
RPV ones (recall $\hat{H}_{d} \equiv \hat{L}_0$). Note that 
$\tilde{L}^\dagger \tilde{m}_{\!\scriptscriptstyle \tilde{L}}^2  \tilde{L}$,
unlike the other soft mass terms, is given by a 
$4\times 4$ matrix. Explicitly, 
$\tilde{m}_{\!\scriptscriptstyle {L}_{00}}^2$ corresponds to 
$\tilde{m}_{\!\scriptscriptstyle H_{\!\scriptscriptstyle d}}^2$ 
of the MSSM case while 
$\tilde{m}_{\!\scriptscriptstyle {L}_{0k}}^2$'s give RPV mass mixings.

The only RPV contribution to the squark masses is given by a
$- (\, \mu_i^*\lambda^{\!\prime}_{ijk}\, ) \; 
\frac{v_{\scriptscriptstyle u}}{\sqrt{2}}$ term in the $LR$ mixing part.
Note that the term contains flavor-changing ($j\ne k$) parts which,
unlike the $A$-terms ones, cannot be suppressed through a flavor-blind
SUSY breaking spectrum. Hence, it has very interesting implications
to quark electric dipole moments (EDMs) and related processses
such as $b\to s\, \gamma$\cite{as4,as6,cch1}.

\thispagestyle{empty}

The mass matrices are a bit more complicated in the scalar sectors\cite{as5,as7}.
We illustrated explicitly here only the charged scalare mass matrix.
The $1+4+3$ charged scalar masses are given in terms of the blocks
\small\beqa
&& \widetilde{\cal M}_{\!\scriptscriptstyle H\!u}^2 =
\tilde{m}_{\!\scriptscriptstyle H_{\!\scriptscriptstyle u}}^2
+ \mu_{\!\scriptscriptstyle \za}^* \mu_{\scriptscriptstyle \za}^{}
+ M_{\!\scriptscriptstyle Z}^2\, \cos\!2 \beta 
\left[ \,\frac{1}{2} - \sin\!^2\theta_{\!\scriptscriptstyle W}\right]
+ M_{\!\scriptscriptstyle Z}^2\,  \sin\!^2 \beta \;
[1 - \sin\!^2 \theta_{\!\scriptscriptstyle W}]
\; ,
\nonumber \\
&&\widetilde{\cal M}_{\!\scriptscriptstyle LL}^2
= \tilde{m}_{\!\scriptscriptstyle {L}}^2 +
m_{\!\scriptscriptstyle L}^\dag m_{\!\scriptscriptstyle L}^{}
+ M_{\!\scriptscriptstyle Z}^2\, \cos\!2 \beta 
\left[ -\frac{1}{2} +  \sin\!^2 \theta_{\!\scriptscriptstyle W}\right] 
+ \left( \begin{array}{cc}
 M_{\!\scriptscriptstyle Z}^2\,  \cos\!^2 \beta \;
[1 - \sin\!^2 \theta_{\!\scriptscriptstyle W}] 
& \quad 0_{\scriptscriptstyle 1 \times 3} \quad \\
0_{\scriptscriptstyle 3 \times 1} & 0_{\scriptscriptstyle 3 \times 3}  
\end{array} \right) 
+ (\mu_{\!\scriptscriptstyle \za}^* \mu_{\scriptscriptstyle \zb}^{})
\; ,
\nonumber \\
&& \widetilde{\cal M}_{\!\scriptscriptstyle RR}^2 =
\tilde{m}_{\!\scriptscriptstyle {E}}^2 +
m_{\!\scriptscriptstyle E}^{} m_{\!\scriptscriptstyle E}^\dag
+ M_{\!\scriptscriptstyle Z}^2\, \cos\!2 \beta 
\left[  - \sin\!^2 \theta_{\!\scriptscriptstyle W}\right] \; ; \qquad
\eeqa
{\normalsize and}
\beqa 
\label{ELH}
\widetilde{\cal M}_{\!\scriptscriptstyle LH}^2
&=& (B_{\za}^*)  
+ \left( \begin{array}{c} 
{1 \over 2} \,
M_{\!\scriptscriptstyle Z}^2\,  \sin\!2 \beta \,
[1 - \sin\!^2 \theta_{\!\scriptscriptstyle W}]  \\
0_{\scriptscriptstyle 3 \times 1} 
\end{array} \right)\; ,
\qquad
\\
\label{ERH}
\widetilde{\cal M}_{\!\scriptscriptstyle RH}^2
&=&  -\,(\, \mu_i^*\lambda_{i{\scriptscriptstyle 0}k}\, ) \; 
\frac{v_{\scriptscriptstyle 0}}{\sqrt{2}} \; ,
\\ 
\label{ERL}
(\widetilde{\cal M}_{\!\scriptscriptstyle RL}^{2})^{\scriptscriptstyle T} 
&=& \left(\begin{array}{c} 
0  \\   A^{\!{\scriptscriptstyle E}} 
\end{array}\right)
 \frac{v_{\scriptscriptstyle 0}}{\sqrt{2}}
-\,(\, \mu_{\scriptscriptstyle \za}^*
\lambda_{{\scriptscriptstyle \za\zb}k}\, ) \, 
\frac{v_{\scriptscriptstyle u}}{\sqrt{2}} \; .
\eeqa \normalsize
Note that $\tilde{m}_{\!\scriptscriptstyle {L}}^2$ here is a $4\times 4$
matrix of soft masses for the $L_\za$, and $B_\za$'s are the corresponding 
bilinear soft terms of the $\mu_{\scriptscriptstyle \za}$'s.
$A^{\!{\scriptscriptstyle E}}$ is just the $3\times 3$ R-parity conserving
leptonic $A$-term. There is no contribution from the admissible RPV $A$-terms
under the SVP. Also, we have used
$m_{\!\scriptscriptstyle L} \equiv \mbox{diag} \{\,0, m_{\!\scriptscriptstyle E}\,\}
\equiv \mbox{diag} \{\,0, m_{\scriptscriptstyle 1}, m_{\scriptscriptstyle 2}, 
m_{\scriptscriptstyle 3}\,\}$.

\section{Neutron Electric Dipole Moment}
Let us take a look first at the quark dipole operator through 1-loop diagrams
with $LR$ squark mixing. A simple direct example is given by the gluino diagram.
Comparing with the MSSM case, the extra (RPV) to the $d$ squark $LR$ mixing
in GSSM obvious modified the story. If
one naively imposes the constraint for this RPV contribution 
itself not to exceed the experimental bound on neutron EDM, one gets roughly
$\mbox{Im}(\mu_i^*\lambda^{\!\prime}_{i\scriptscriptstyle 1\!1}) 
\lsim 10^{-6}\,\mbox{GeV}$, a constraint that is interesting even
in comparison to the bounds on the corresponding parameters obtainable
from asking no neutrino masses to exceed the super-Kamiokande 
atmospheric oscillation scale\cite{as4}. 

In fact, there are important contributions beyond the gluino diagram and without 
$LR$ squark mixings involved. For the MSSM, it is well-known that there is such
a contribution from the chargino diagram, which is likely to be more important
than the gluino one when a unification type gaugino mass relationship is
imposed. The question then is if the GSSM has a similar RPV analog. A RPV
version of the chargino diagram is given in Fig.1. The diagram, however, looks
ambiguous. Looking at the diagram in terms of the electroweak states involved
under our formulation, it seems like a 
${l}_k^{\!\!\mbox{ -}}$--$\tilde{W}^{\scriptscriptstyle +}$
mass insertion is required, which is however vanishing. However, putting in
extra mass insertion, with a $\mu_i$ flipping the ${l}_k^{\!\!\mbox{ -}}$ into 
a $\tilde{h}_u^{\scriptscriptstyle +}$ first seems to give a non-zero result.
The structure obviously indicates a GIM-like cancellation at worked, and we
have to check its violation due to the lack of mass degeneracy.

\begin{figure}[h]
\includegraphics{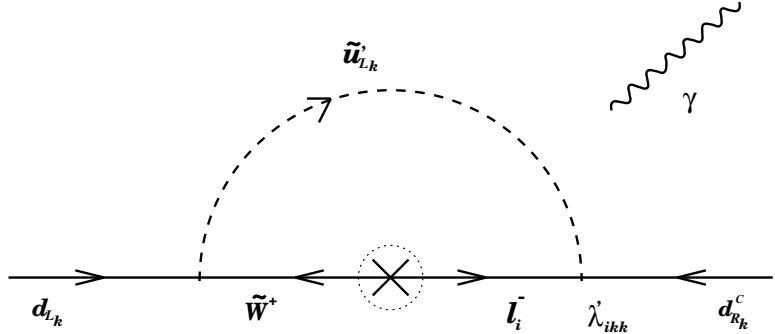}
\vspace*{2in}
\caption{The new charginolike diagram.}
\end{figure}

\thispagestyle{empty}

We have performed an extensive analytical and numerical study, including the 
complete charginolike contributions, as well as the neutralinolike contributions,
to the neutron EDM\cite{as6}. The charginolike part is given by the following
formula :
\beq \label{edmco}
\left({d_{\scriptscriptstyle f} \over e} \right)_{\!\!\chi^{\mbox{-}}} = 
{\alpha_{\!\mbox{\tiny em}} \over 4 \pi \,\sin\!^2\theta_{\!\scriptscriptstyle W}} \; 
\sum_{\scriptscriptstyle \tilde{f}'\mp} 
\sum_{n=1}^{5} \,\mbox{Im}({\cal C}_{\!fn\mp}) \;
{{M}_{\!\scriptscriptstyle \chi^{\mbox{-}}_n} \over 
M_{\!\scriptscriptstyle \tilde{f}'\mp}^2} \;
\left[ {\cal Q}_{\!\tilde{f}'} \; 
B\!\left({{M}_{\!\scriptscriptstyle \chi^{\mbox{-}}_{n}}^2 \over 
M_{\!\scriptscriptstyle \tilde{f}'\mp}^2} \right) 
+ ( {\cal Q}_{\!{f}} - {\cal Q}_{\!\tilde{f}'} ) \;
A\!\left({{M}_{\!\scriptscriptstyle \chi^{\mbox{-}}_{n}}^2 \over 
M_{\!\scriptscriptstyle \tilde{f}'\mp}^2} \right) 
\right] \; ,
\eeq 
for $f$ being $u$ ($d$) quark and $f'$ being $d$ ($u$), where
\beqa
{\cal C}_{un-} &=&  
{y_{\!\scriptscriptstyle u} \over g_{\scriptscriptstyle 2} } \,\, 
\mbox{\boldmath $V$}^{\!*}_{\!\!2n} \, {\cal D}_{d11} \;
\left( - \mbox{\boldmath $U$}_{\!1n} \,{\cal D}^{*}_{d11} 
+ {y_{\!\scriptscriptstyle d} \over g_{\scriptscriptstyle 2} }\,\, 
\mbox{\boldmath $U$}_{\!2n}\,  {\cal D}^{*}_{d21}
+ {\lambda^{\!\prime}_{k\scriptscriptstyle 1\!1} \over g_{\scriptscriptstyle 2} }\,\, 
\mbox{\boldmath $U$}_{\!(k+2)n}\,  {\cal D}^{*}_{d21} \right) \; ,
\nonumber \\
{\cal C}_{un+} &=&
 {y_{\!\scriptscriptstyle u} \over g_{\scriptscriptstyle 2} } \,\, 
\mbox{\boldmath $V$}^{\!*}_{\!\!2n} \, {\cal D}_{d12} \;
\left( - \mbox{\boldmath $U$}_{\!1n} \, {\cal D}^{*}_{d12} 
+ {y_{\!\scriptscriptstyle d} \over g_{\scriptscriptstyle 2} }\,\, 
\mbox{\boldmath $U$}_{\!2n}\,  {\cal D}^{*}_{d22}
+ {\lambda^{\!\prime}_{k\scriptscriptstyle 1\!1} \over g_{\scriptscriptstyle 2} }\,\, 
\mbox{\boldmath $U$}_{\!(k+2)n}\,  {\cal D}^{*}_{d22} \right) \; ,
\nonumber \\
{\cal C}_{dn-} &=& 
\left( {y_{\!\scriptscriptstyle d} \over g_{\scriptscriptstyle 2} }\,\, 
\mbox{\boldmath $U$}_{\!2n} 
+ {\lambda^{\!\prime}_{k\scriptscriptstyle 1\!1} \over g_{\scriptscriptstyle 2} }\,\, 
\mbox{\boldmath $U$}_{\!(k+2)n} \right)\! {\cal D}_{u11} \;
\left( - \mbox{\boldmath $V$}^{\!*}_{\!\!1n} \,{\cal D}^{*}_{u11} 
+ {y_{\!\scriptscriptstyle u} \over g_{\scriptscriptstyle 2} } \,
\mbox{\boldmath $V$}^{\!*}_{\!\!2n} \, {\cal D}^{*}_{u21} \right) \; ,
\nonumber \\
{\cal C}_{dn+} &=& 
\left( {y_{\!\scriptscriptstyle d} \over g_{\scriptscriptstyle 2} }\,\, 
\mbox{\boldmath $U$}_{\!2n} 
+ {\lambda^{\!\prime}_{k\scriptscriptstyle 1\!1} \over g_{\scriptscriptstyle 2} }\,\, 
\mbox{\boldmath $U$}_{\!(k+2)n} \right)\! {\cal D}_{u12} \;
\left( - \mbox{\boldmath $V$}^{\!*}_{\!\!1n} \, {\cal D}^{*}_{u12} 
+ {y_{\!\scriptscriptstyle u} \over g_{\scriptscriptstyle 2} } \,
 \mbox{\boldmath $V$}^{\!*}_{\!\!2n} \, {\cal D}^{*}_{u22} \right) \; ,
\nonumber \\
&& \mbox{\hspace*{2.5in}\small(only repeated index $i$ is to be summed)} \; ;
\label{Cnmp}
\eeqa
$\mbox{\boldmath $V$}^\dag {\mathcal{M}_{\scriptscriptstyle C}} \,
\mbox{\boldmath $U$} = \mbox{diag} 
\{ {M}_{\!\scriptscriptstyle \chi^{\mbox{-}}_n} \} \equiv 
\mbox{diag} 
\{ {M}_{c {\scriptscriptstyle 1}}, {M}_{c {\scriptscriptstyle 2}},
m_e, m_\mu, m_\tau \}$ while ${\cal D}_{u}$ and ${\cal D}_{d}$ diagonalize
the $\tilde{u}$ and $\tilde{d}$ squark mass-squared matrices respectively;
and
\beq
A(x) = {1 \over 2 \, (1-x)^2} \left(3 - x + {2\ln x \over 1-x} \right)\;,
\qquad \qquad
B(x) = {1 \over 2\,(x-1)^2} \left[1 + x + {2\,x \ln x \over (1-x) } \right]. 
\nonumber 
\eeq

To extract the contribution from the diagram of Fig.~1, we have to look at the
pieces in ${\cal C}_{dn\mp}$ with a $\mbox{\boldmath $V$}^{\!*}_{\!\!1n}$
and a $\mbox{\boldmath $U$}_{\!(k+2)n}$. It is easy to see that the $n=1$ and 
$2$ mass eigenstates, namely the chargino states, do give the dominating
contribution. With the small $\mu_i$ mixings strongly favored by the 
sub-eV neutrino masses, we have 
\beq
\mbox{\boldmath $U$}_{\!(k+2)1} =
\frac{{\mu_k^*}}{{M}_{c {\scriptscriptstyle 1}}}
R_{\!\scriptscriptstyle R_{21}}  
\qquad \mbox{and} \qquad
\mbox{\boldmath $U$}_{\!(k+2)2} =
\frac{{\mu_k^*}}{{M}_{c {\scriptscriptstyle 2}}}
R_{\!\scriptscriptstyle R_{22}} 
\eeq
where the $R_{\!\scriptscriptstyle R}$ denotes the right-handed rotation 
that would diagonalize the first $2\times 2$ block of 
${\mathcal{M}_{\scriptscriptstyle C}}$. The latter rotation matrix is expected
to have elements of order 1. Hence, we have the dominating result
proportional to 
\[ 
\sum_{n=1,2}
R_{\!\scriptscriptstyle R_{12}}^{\,*} \,
R_{\!\scriptscriptstyle R_{2}n} \;
{\mu_k^* \,\lambda^{\!\prime}_{k\scriptscriptstyle 1\!1}} \;
F_{\!\scriptscriptstyle B\!A}\!\!
\left( M_{c_{\scriptscriptstyle n}}^2 \right) 
\]
where $F_{\!\scriptscriptstyle B\!A}$ denotes the mass eigenvalue
dependent part. The result agrees with what we say above. It vanishes for
${M}_{c {\scriptscriptstyle 1}}={M}_{c {\scriptscriptstyle 2}}$, showing
a GIM-like mechanism. However, with unequal chargino masses, our numerical
results indicate that the cancellation is generically badly violated.
More interestingly, it can be seen from the above analysis that a complex
phase in ${\mu_k^* \,\lambda^{\!\prime}_{k\scriptscriptstyle 1\!1}}$
is actually no necessary for this potentially dominating chargino 
contribution to be there, so long as complex CP violating phases exist
in the $R_{\!\scriptscriptstyle R}$ matrix, {\it i.e.} in the R-parity
conserving parameters such as ${\mu_{\scriptscriptstyle 0}}$.

An illustration of the result is given in Fig.~3 in which variations of the 
EDM contribution against the $\tan\!\zb$ value is plotted.
On the whole, the magnitude of the parameter combination
$\mu_i^*\lambda^{\!\prime}_{i\scriptscriptstyle 1\!1}$ is shown to be 
responsible for the RPV 1-loop contribution to neutron EDM and is hence 
well constrained. This applies not only to the complex phase, or imaginary part 
of, the combination. Readers are referred to Ref.\cite{as6} for more details.

\thispagestyle{empty}

\begin{minipage}[b]{\textwidth}
\twocolumn

\begin{figure}[b]
\vspace*{10.5cm}
\includegraphics{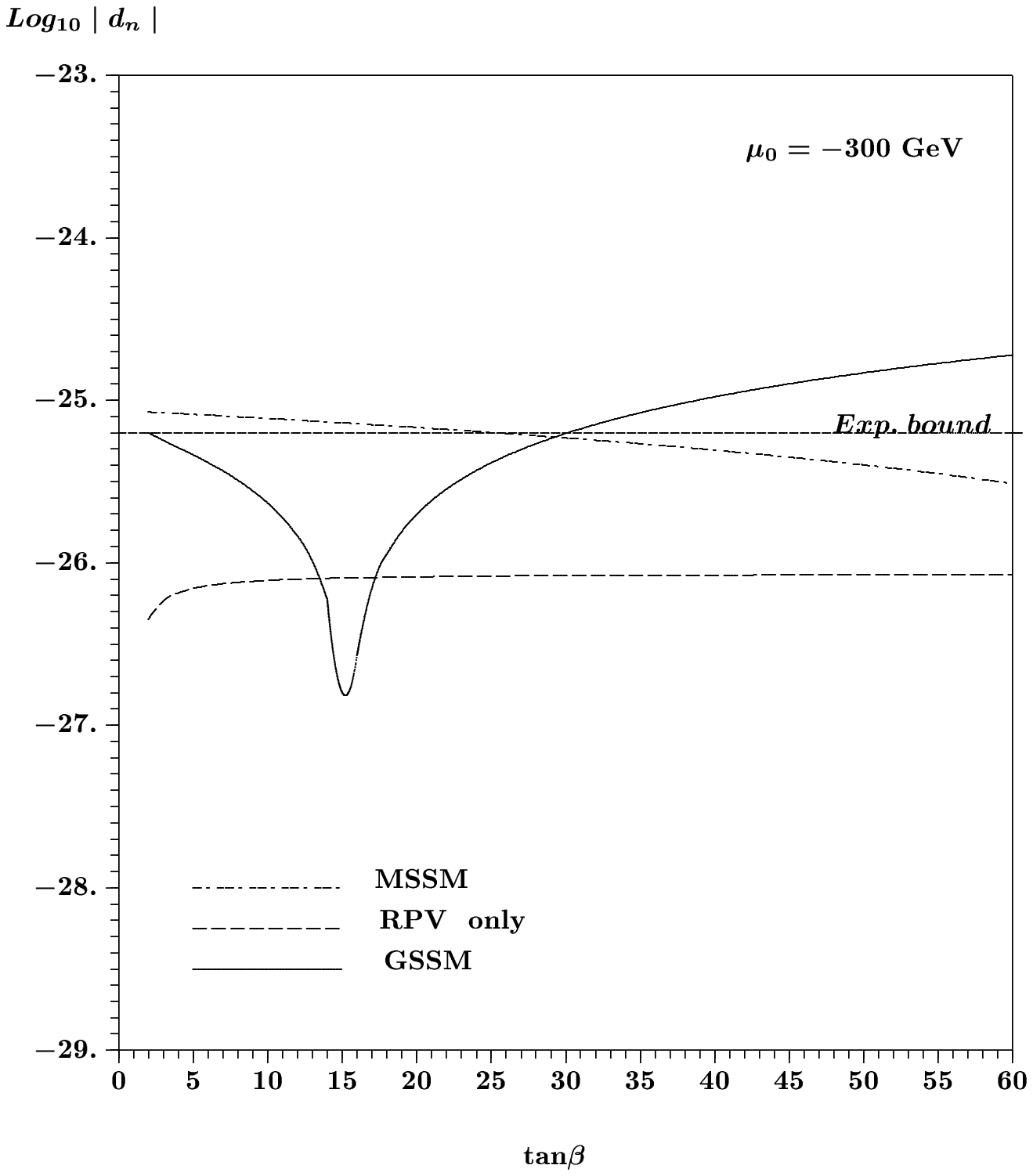}
\end{figure}
\vspace*{-.1in}
\parbox{7.5cm}{\small FIG.2
Logarithmic plot of (the magnitude of) the neutron EDM result verses 
$\tan\!\zb$. We show here the MSSM result, our general result with RPV phase
only, and the generic result  with complex phases of 
both kinds. In particular, the $A$ and $\mu_{\scriptscriptstyle 0}$ 
phases are chosen as $7^o$ and $0.1^o$ respectively, for the MSSM line. They are 
zero for the RPV-only line, with which we have a phase of ${\pi\over 4}$ for 
$\lambda^{\!\prime}_{\scriptscriptstyle 31\!1}$. All the given nonzero values
are used for the three phases for the generic result (from our complete formulae) 
marked by GSSM. 
\vspace*{.5in}}
\end{minipage}

\begin{minipage}[b]{\textwidth}
\vspace*{7.5cm}
\parbox{7.3cm}
{ 
\section{Dipole Moments of the Electron and Other Fermions}
\hspace*{0.5cm}
There is in fact a second class of 1-loop diagrams contributing to the
quark EDMs. These are diagrams with quarks and scalars in the loop, and
hence superpartners of the charginolike and neutralinolike diagrams discussed
above. The R-parity
conserving analog of the class of diagrams has no significance, due to
the unavoidable small Yukawa couplings involved. With the latter replaced
by flavor-changing $\lambda^{\!\prime}$-couplings. We can have a $t$ quark
loop contributing to neutron EDM, for example. 

\hspace*{0.5cm}
For the case of the
charged leptons, the two classes of superpartner diagrams merges into one.
But then, all scalars has to be included. The assumption hidden, in our
quark EDM formula above, that only the (two) superpartner sfermions
have a significant role to play does not stand any more. 

\hspace*{0.5cm} 
The above quark EDM formula obviously applies with some trivial modifications
to the cases of the other quarks. For the charge leptons, while the exact formulae
would be different, there are major basic features that are more or less
the same. For instance, for the charged lepton, the $\lambda$-couplings
play the role of the $\lambda^{\!\prime}$-couplings. The 
$\mu_i^*\lambda_{i\scriptscriptstyle 1\!1}$ combination contributes to 
electron EDM while the $\mu_i^*\lambda_{i\scriptscriptstyle 22}$ 
combination contributes to that of the muon. As we have no explicit 
numerical results to show at the moment, we refrain from showing any details
here. However, we have finished a $\mu \to  e \,\gamma$ 
study\cite{as7}, 

}\end{minipage}
\onecolumn 

\noindent
from which the charged lepton EDM 
formula could be extracted without too much effort.
Interested readers may check the reference for details.

\thispagestyle{empty}

\section{Neutrino dipole moments}
Another topic we want to discuss briefly here
is the dipole moments of the neutrinos. Neutrinos as 
Majorana fermions have vanishing dipole moments. However, flavor
off-diagonal dipole moments, or known as transition dipole moments
are interesting. There are good terrestial as well as astrophysical
and cosmological bounds available\cite{nudm}. 

The same set of diagrams giving rise to 1-loop neutrino masses
within the model give rise also to dipole moments when an extra
photon line is attached. There are two types of such neutrino
mass diagrams, the charged and neutral loop ones. A neutral
loop diagram has, of course, no place to attach a photon line.
Hence, only the charged loop diagrams contribute.  Checking
parameter fits to both neutrino masses and their implications
on dipole moments would be very interesting.

We give in Ref.\cite{as9}, all contributions
to 1-loop neutrino masses within GSSM under a systematic
framework. For example, each diagram composes of
two (external) neutrino interaction vertices. The charged
vertices are given by 
\beqa
{\cal C}^{\scriptscriptstyle R}_{inm} 
&=&
\frac{y_{\!\scriptscriptstyle e_i}}{g_{\scriptscriptstyle 2}} \,
\mbox{\boldmath $V$}_{\!\!(i+2)n} \, D^{l^*}_{\!2m} 
- \frac{\lambda_{ikh}^{\!*}}{g_{\scriptscriptstyle 2}} 
\mbox{\boldmath $V$}_{\!\!(h+2)n} \, D^{l^*}_{\!(k+2)m}  \; ,
\nonumber \\
{\cal C}^{\scriptscriptstyle L}_{inm} 
&=&  
- \mbox{\boldmath $U$}_{\!1n} \, {\cal D}^{l^*}_{\!(i+2)m}
+ \frac{y_{\!\scriptscriptstyle e_i}}{g_{\scriptscriptstyle 2}} \,
  \mbox{\boldmath $U$}_{\!2n} \, {\cal D}^{l^*}_{\!(i+5)m}
- {\lambda_{ihk} \over g_{\scriptscriptstyle 2} } \, 
\mbox{\boldmath $U$}_{\!(h+2)n} \, {\cal D}^{l^*}_{\!(k+5)m} \; .
\label{Cnm}
\eeqa
A ${\cal C}^{\scriptscriptstyle R^*}_{jnm} $ 
${\cal C}^{\scriptscriptstyle L}_{inm} $ combination plays the role
of ${\cal C}_{\!fn\mp}$ in the formula of Eq.(\ref{edmco}), for $\nu_i$
and $\nu_j$. Here, we are interested not only in the imaginary part; 
the real part contribute magnetic moments. Nevertheless, we have to
switch back to the mass eigenstate basis for the neutrinos to better
understand and use the dipole moment results\cite{new}.

\end{document}